\documentclass[twocolumn,english,aps,pra,reprint,superscriptaddress,nofootinbib]{revtex4-2}
\usepackage[T1]{fontenc}
\usepackage[latin9]{inputenc}
\usepackage{amsmath}
\usepackage{amssymb}
\usepackage{graphicx}
\usepackage{wasysym}
\usepackage{color}
\usepackage{microtype}
\usepackage{mhchem}

\makeatletter

\usepackage{pslatex}

\usepackage{babel}

\makeatother

\usepackage{hyperref}

\usepackage{xcolor}
\hypersetup{
	colorlinks,
	linkcolor={blue!70!black},
	citecolor={blue!70!black},
	urlcolor={blue!70!black}
}

\begin{document}

\title{Blackbody thermalization and vibrational lifetimes of trapped polyatomic molecules}

\author{Nathaniel B. Vilas}
\email{vilas@g.harvard.edu}
\affiliation{Department of Physics, Harvard University, Cambridge, MA 02138, USA}
\affiliation{Harvard-MIT Center for Ultracold Atoms, Cambridge, MA 02138, USA}

\author{Christian Hallas}
\affiliation{Department of Physics, Harvard University, Cambridge, MA 02138, USA}
\affiliation{Harvard-MIT Center for Ultracold Atoms, Cambridge, MA 02138, USA}

\author{Lo\"ic Anderegg}
\affiliation{Department of Physics, Harvard University, Cambridge, MA 02138, USA}
\affiliation{Harvard-MIT Center for Ultracold Atoms, Cambridge, MA 02138, USA}

\author{Paige Robichaud}
\affiliation{Department of Physics, Harvard University, Cambridge, MA 02138, USA}
\affiliation{Harvard-MIT Center for Ultracold Atoms, Cambridge, MA 02138, USA}

\author{Chaoqun Zhang}
\affiliation{Department of Chemistry, The Johns Hopkins University, Baltimore, MD 21218, USA}

\author{Sam Dawley}
\affiliation{Department of Chemistry, The Johns Hopkins University, Baltimore, MD 21218, USA}

\author{Lan Cheng}
\affiliation{Department of Chemistry, The Johns Hopkins University, Baltimore, MD 21218, USA}

\author{John M. Doyle}
\affiliation{Department of Physics, Harvard University, Cambridge, MA 02138, USA}
\affiliation{Harvard-MIT Center for Ultracold Atoms, Cambridge, MA 02138, USA}

\date{\today}

\begin{abstract}

We study the internal state dynamics of optically trapped polyatomic molecules subject to room temperature blackbody radiation. Using rate equations that account for radiative decay and blackbody excitation between rovibrational levels of the electronic ground state, we model the microscopic behavior of the molecules' thermalization with their environment. As an application of the model, we describe in detail the procedure used to determine the blackbody and radiative lifetimes of low-lying vibrational states in ultracold CaOH molecules, the values of which were reported in previous work [Hallas et al., arXiv:2208.13762]. \emph{Ab initio} calculations are performed and are found to agree with the measured values. Vibrational state lifetimes for several other laser-coolable molecules, including SrOH and YbOH, are also calculated.

\end{abstract}

\maketitle

\section{Introduction}

Cold and ultracold polyatomic molecules are a promising resource for diverse applications that span quantum information science \cite{wei2011entanglement, yu2019scalable}, quantum simulation \cite{wall2013simulating,wall2015quantum,wall2015realizing}, ultracold collisions \cite{augustovivcova2019ultracold}, cold chemistry \cite{heazlewood2021towards}, and searches for physics beyond the standard model \cite{kozyryev2017precision,kozyryev2021enhanced,hutzler2020polyatomic}. While ultracold diatomic molecules are now routinely created and studied in the laboratory \cite{de2019degenerate, anderegg2019optical, schindewolf2022evaporation, christakis2022probing, holland2022on, bao2022dipolar}, ultracold polyatomic molecules, with their increased vibrational and rotational degrees of freedom, have only more recently begun to be brought under single quantum state control. In the last decade, CH$_3$F \cite{zeppenfeld2012sisyphus, koller2022electric} and H$_2$CO \cite{prehn2016optoelectrical} have been cooled and trapped via optoelectrical Sisyphus cooling, CH$_3$ has been magnetically trapped \cite{liu2017magnetic}, and CaOH has been laser cooled and trapped in a magneto-optical trap (MOT) \cite{vilas2022magneto}, then loaded into an optical trap \cite{hallas2022optical}. SrOH, YbOH, and CaOCH$_3$ have been laser cooled in one dimension \cite{kozyryev2017sisyphus,augenbraun2019laser,baum20201d,mitra2020direct}, setting the stage for MOTs and optical traps of an increasing variety of polyatomic molecules.

One challenge of working with trapped polar molecules is their susceptibility to loss from blackbody radiation from the environment, which incoherently drives rovibrational transitions out of the internal quantum state of interest. This mechanism has been theoretically studied \cite{vanhaecke2007precision, buhmann2008surface, norrgard2022quantum} and experimentally observed \cite{hoekstra2007optical, williams2018magnetic, chou2020frequency, boon2022spectroscopy} for trapped diatomic molecules, where it can limit lifetimes to $\lesssim10$~s at room temperature. The specific sensitivity of diatomic molecules to blackbody radiation has also been proposed as a tool for the calibration of temperature standards~\cite{norrgard2022quantum}. Because of the increased quantity of rovibrational states in polyatomic molecules (the number of vibrational modes scales as $\sim3N$, where $N$ is the number of atoms in the molecule), they have the potential for significantly increased sensitivity to blackbody radiation compared to diatomic species. Although blackbody lifetimes of trapped polyatomic molecules have received some attention in the literature \cite{glockner2015rotational, liu2017magnetic, zeppenfeldthesis}, they have been less thoroughly explored.

In addition to blackbody loss, excited vibrational states in polar molecules have finite lifetimes due to spontaneous, radiative decay. A number of proposed experiments with polyatomic molecules rely on population in excited vibrational levels, and the radiative lifetimes of these states are therefore critical for these experiments. In particular, the low-lying vibrational bending modes of linear polyatomic molecules like CaOH, SrOH, and YbOH are expected to be useful for experiments including quantum simulation and computation \cite{wall2013simulating, wall2015realizing, yu2019scalable}, electrostatic shielding of collisions \cite{augustovivcova2019ultracold}, and precision searches for ultralight dark matter and the electric dipole moment of the electron \cite{kozyryev2021enhanced, kozyryev2017precision, anderegg2023quantum}. Radiative decay lifetimes for excited vibrational states have been previously measured in diatomic molecules \cite{campbell2008time, versolato2013decay, boon2022spectroscopy}, but these measurements are more challenging in polyatomic molecules due to their increasingly complex vibrational structure.

In this work, we study the vibrational state lifetimes of polyatomic molecules resulting from blackbody excitation and radiative decay. The combination of these effects drives the molecules into equilibrium with their thermal environment. Using a rate equation model and \emph{ab initio} calculations of transition rates, we explore the evolution of rovibrational state populations over time and compute blackbody and radiative decay lifetimes for laser-coolable, linear triatomic molecules. We describe a fit of the rate equations to experimental data for optically trapped CaOH, which was used to determine the radiative and blackbody lifetimes reported in Ref. \cite{hallas2022optical}. The fit results are compared with \emph{ab initio} calculations. Finally, we perform calculations of ground-state blackbody lifetimes for larger laser-coolable polyatomic molecules, which have additional vibrational degrees of freedom susceptible to blackbody loss.

The structure of this paper is as follows. In section \ref{sec:RateEquations}, we describe a set of rate equations used to model blackbody thermalization in trapped polyatomic molecules. In section \ref{sec:caoh}, we detail the application of this model to measure the blackbody and radiative lifetimes of low-lying vibrational levels in optically trapped CaOH molecules \cite{hallas2022optical}. In section \ref{sec:abinitio}, we describe \emph{ab initio} calculations of radiative and blackbody lifetimes for CaOH, SrOH, and YbOH molecules. In section \ref{sec:larger}, we discuss calculated blackbody lifetimes for larger and more asymmetric polyatomic molecules, which can have many more vibrational modes. In section \ref{sec:conclusion}, we conclude.

\section{Rate equation model}
\label{sec:RateEquations}

\begin{figure}
    \centering
    \includegraphics{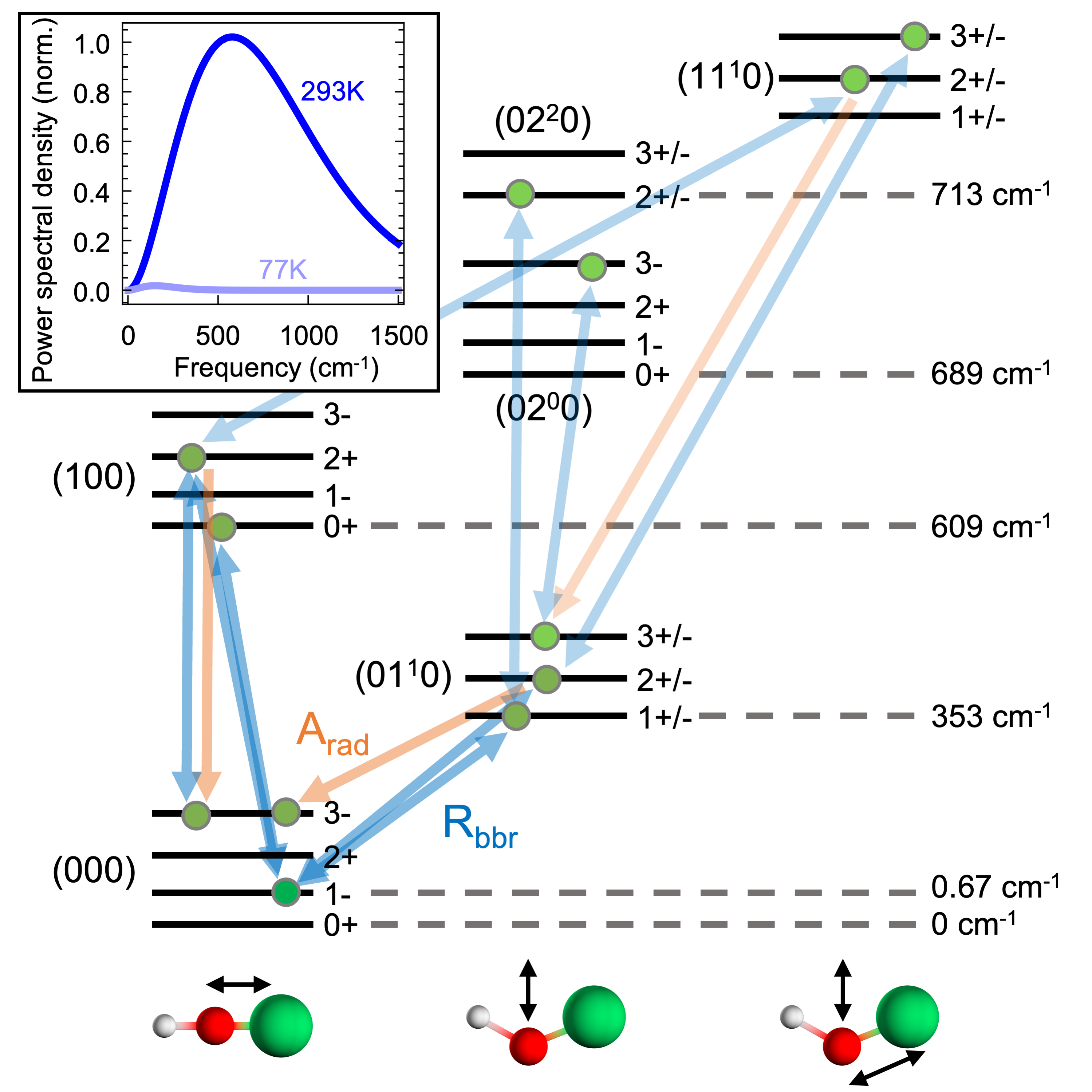}
    \caption{Schematic level diagram illustrating rovibrational thermalization dynamics in low-lying vibrational levels of CaOH, with all population initially prepared in the $N=1$ level of the $(000)$ state. Populations are represented by green circles, while blackbody transitions and radiative decay are represented by blue and orange arrows, respectively. For clarity, only some representative transitions are shown. All rovibrational transitions are driven following the selection rules (within the harmonic approximation) $\Delta v_1 = \pm 1$ or $\Delta v_2 = \pm 1$, $\Delta N = 0,\pm 1$, and $\Delta p = \pm 1$. Transition rates for the blackbody transitions, $R_\text{bbr}$, are determined by the power spectral density of blackbody radiation at the transition frequency (inset). The antisymmetric (\ce{O-H}) stretching mode in CaOH has a frequency of $\sim$3700 cm$^{-1}$ \cite{li1996dye} and is therefore neglected.}
    \label{fig:leveldiagram}
\end{figure}

In typical experiments with cold, trapped molecules, the molecules are prepared in a single internal quantum state, embedded in a large landscape of states arising from hyperfine, rotational, vibrational, and electronic structure. This highly nonthermal initial distribution can be affected by the room temperature environment. The thermalization process, during which the internal state distribution of the molecule comes into thermal equilibrium with the environment, is mediated by blackbody radiation. Microscopically, thermalization occurs when the tendency for blackbody radiation to drive transitions up the rovibrational ladders is balanced by radiative decay, as first described by the $A$ and $B$ rate coefficients of Einstein.

In polar molecules, electric dipole transitions between rotational or vibrational levels can be strongly driven by room temperature blackbody radiation, which has a maximum power spectral density near 600 cm$^{-1}$. Because vibrational frequencies in laser coolable molecules are often near this energy, blackbody-driven vibrational transitions can be a significant source of incoherent population transfer in trapped molecules on experimentally relevant timescales, typically on the order of 1 s \cite{hoekstra2007optical, williams2018magnetic, chou2020frequency}.

Fig. \ref{fig:leveldiagram} schematically illustrates these dynamics using CaOH molecules as an example. The \ce{Ca-O} stretching mode and the bending mode both have vibrational frequencies near the peak of the blackbody spectrum at 300K (inset). Blackbody radiation drives transitions at a rate $R_\text{bbr}$ between vibrational states obeying the electric dipole selection rules $\Delta v_1 = \pm1$ or $\Delta v_2 = \pm1$; $\Delta \ell = 0,\pm1$; $\Delta N = 0,\pm1$; and $\Delta p = \pm 1$, where $v_1$ and $v_2$ are the vibrational quantum numbers for the Ca--O stretch and \ce{Ca-O-H} bending modes, $\ell$ is the projection of the vibrational angular momentum onto the molecular axis, $N$ is the rotational quantum number, and $p$ is the parity of the state. Excited states spontaneously decay according to the same selection rules at a rate $A_\text{rad}$.

In typical laser cooling experiments, the population is initially prepared in the $N=1$ level of the vibrational ground state, as shown in Fig. \ref{fig:leveldiagram}. Vibrational populations thermalize after a few transitions according to the selection rules highlighted above, while rotational populations take many transitions to fully thermalize, as the populations ``walk'' up the rotational ladder via repeated transitions between vibrational states. The timescale for rotational thermalization is therefore significantly longer than for vibrational thermalization. (Note that pure rotational transitions are driven by the blackbody environment much more slowly due to their low energy.) In this section, we describe a rate equation model to quantitatively describe these dynamics.

\subsection{Thermalization dynamics}

We model the internal state dynamics of conservatively trapped polyatomic molecules using a set of rate equations that capture the effects of radiative decay and blackbody excitation between vibrational manifolds within the electronic ground state. We consider rovibronic states described by the quantum numbers $\lvert \{v\}_i, N, K, p\rangle$, where $\{v\}_i \equiv \{v_{1i}, v_{2i}, \ldots \}$ is the set of quantum numbers describing excitation of the vibrational normal modes, $N$ is the rotational quantum number, $K$ is the projection of $N$ onto the molecule-frame $z$ axis, and $p$ is the parity of the state. Spin-rotation and hyperfine structure can be readily included in the equations below using standard angular momentum algebra, but they do not affect the primary results of this section.

The spontaneous decay rate, $A_{ij}$, and the blackbody excitation rate, $R_{ij}$, from initial state $i$ to final state $j$ are \cite{bernath2005spectra}
\begin{align}
    A_{ij} &= \frac{\omega_{ij}^3}{3\pi\varepsilon_0\hbar c^3 (2N_i +1)}S_{ij}
    \label{eqn:spont},
    \\
    R_{ij} &= \frac{1}{6\varepsilon_0\hbar^2(2N_i+1)}\frac{2\hbar \omega_{ij}^3}{\pi c^3}\frac{1}{e^{\hbar\omega_{ij}/k_BT}-1}S_{ij}
    \label{eqn:BBR},
\end{align}
where $S_{ij} = \left\lvert \langle i \lvert \mu \lvert j \rangle \right\lvert^2$ is the rovibrational transition strength, $N_i$ is the angular momentum quantum number of the initial state, $\omega_{ij}$ is the transition frequency, and $T$ is the temperature of the environment.

\begin{figure*}
    \centering
    \includegraphics{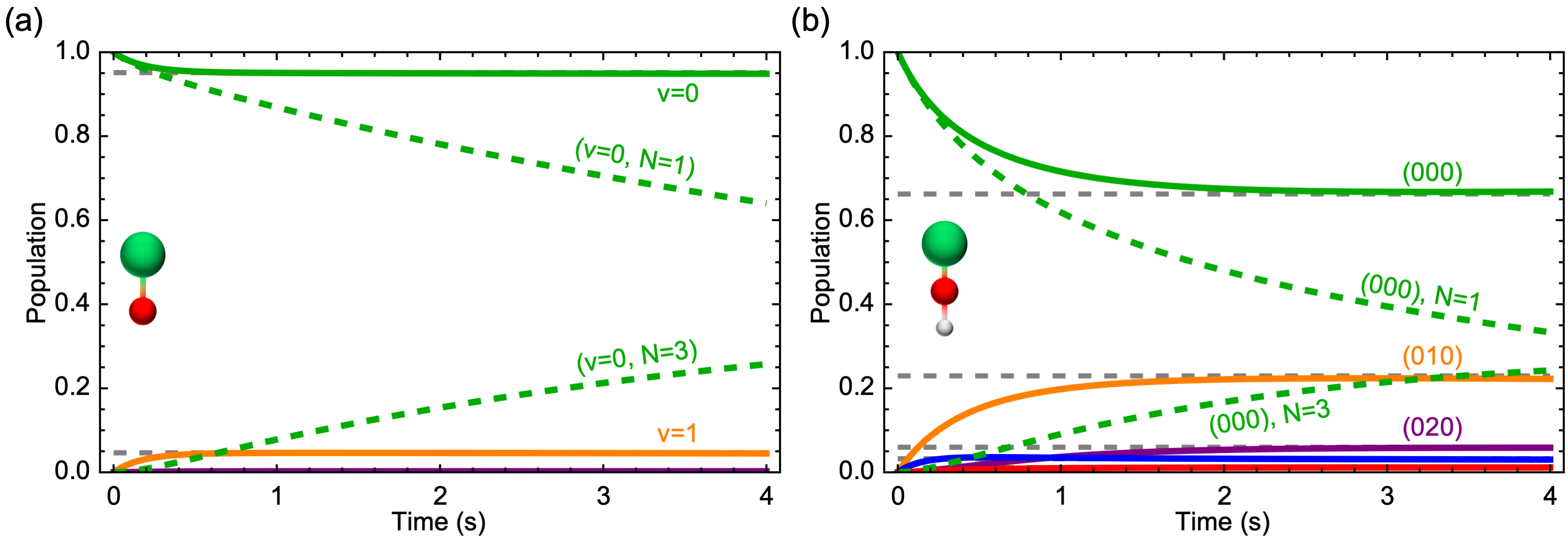}
    \caption{Generalized calculations of vibrational thermalization dynamics for (a) diatomic molecules and (b) linear triatomic molecules. For these plots, the molecular constants are matched to those of CaOH, but with the bending vibration removed in the diatomic case (resulting in a molecule similar to CaF). The inclusion of one additional vibrational mode significantly decreases the lifetime of the ground state $N=1$ level and increases the time required to achieve vibrational thermalization. Dashed gray lines correspond to the equilibrium vibrational populations.}
    \label{fig:rateeqnplots}
\end{figure*}

The rovibrational transition strengths can be separated into vibrational and rotational components:
\begin{align}
    S_{ij} &= \left\lvert \langle \{v\}_i, N_i, K_i, p_i \lvert \mu \lvert \{v\}_j, N_j, K_j, p_j \rangle \right\lvert^2 \nonumber \\
    &= \left\lvert \langle \{v\}_i \lvert \mu \lvert \{v\}_j \rangle \right\lvert^2 \left\lvert \langle N_i, K_i, p_i \lvert \mu \lvert N_j, K_j, p_j \rangle \right\lvert^2 \nonumber \\
    &\equiv S^\text{vib}_{ij} S^\text{rot}_{ij},
\end{align}
where $S^\text{rot}_{ij}$ and $S^\text{vib}_{ij}$ are the rotational and vibrational line strengths, respectively. The rotational line strengths, or H\"onl-London factors, are \cite{hansson2005comment, hirota2012high}
\begin{align}
    S^\text{rot}_{ij} = &\delta_{p_i,-p_j}(1+\delta_{K_i 0} + \delta_{K_j 0} - 2 \delta_{K_i0}\delta_{K_j0})\nonumber \\
    & \times (2N_i+1)(2N_j+1)
    \begin{pmatrix}
    N_i & 1 & N_j \\
    -K_i & K_i-K_j & K_j
    \end{pmatrix}
    ^2
\label{eqn:hlvib},
\end{align}
where we emphasize that since we have chosen a basis where the parity is defined for each state, $K$ is always a positive number. The Kronecker delta terms enforce a 2$\times$ increase in the line strength for transitions from nondegenerate ($K=0$) to degenerate ($K\neq 0$) states \cite{whiting1980recommended}.

For the vibrational line strengths, we make the ``double harmonic'' approximation \cite{bernath2005spectra}, wherein the vibrational potential is assumed to be that of a perfect harmonic oscillator, and the electric dipole moment is assumed to be a linear function of the internuclear spacing near the equilibrium geometry of the molecule. Within this approximation, we expand the vibrational line strength as follows:
\begin{align}
    \langle \{v\}_i \lvert \mu \lvert \{v\}_j \rangle \approx \sum_n^{3N-6(5)}\left|\frac{d\vec{\mu}_e}{dQ_n}\right|_{Q_{n,\text{eq}}} \langle v_{n,i}\lvert Q_n \lvert v_{n,j}\rangle
    \label{eqn:dipoleexpansion},
\end{align}
where the sum is over all $3N-6$ ($3N-5$ in a linear molecule) vibrational modes, $\vec{\mu}_e$ is the electric dipole moment of the molecule, $Q_n$ is a normal coordinate describing the $n$th mode, and $Q_{n,\text{eq}}$ is its equilibrium value.\footnote{Our units are chosen so that $d\vec{\mu}_e/dQ_n$ has units of debye; this can be achieved by making $Q_n$ unitless by scaling by the harmonic oscillator length $\sqrt{\hbar/(\mu\omega_n)}$, where $\mu$ is the reduced mass of the vibration.}  Within the harmonic approximation, we can use harmonic oscillator algebra to write the vibrational matrix elements. For nondegenerate vibrational modes the matrix elements are
\begin{align}
    \left\lvert\langle v_n+1|Q_n|v_n\rangle\right\lvert^2 &= \frac{v_n+1}{2},
\end{align}
while for degenerate modes we use higher-dimensional harmonic oscillator algebra. For instance, for vibrational bending modes with vibrational angular momentum $\ell$ (as found in, e.g., linear triatomic molecules), the degeneracy is 2 and the matrix elements are
\begin{align}
    &\left\lvert\langle v_n+1, \ell\pm1|Q_n|v_n,\ell\rangle\right\lvert^2 \nonumber \\
    &= \frac{1}{4}(1+ \delta_{\ell,0} + \delta_{\ell\pm1,0}-\delta_{\ell,0}\delta_{\ell\pm1,0})\left(\frac{v_2\pm\ell}{2}+1\right).
\end{align}

The advantage of the double-harmonic approximation is that it allows every vibrational transition moment to be expressed in terms of just the dipole moment derivatives along each of the normal coordinates. For low-lying vibrational states this approximation is expected to be reasonably appropriate. If deemed necessary, anharmonicity can be included by expanding eqn. \ref{eqn:dipoleexpansion} to higher order. The matrix elements $\langle v_{n,i} \lvert Q_n^k \lvert v_{n,j}\rangle$ can be calculated using standard harmonic oscillator algebra and will be able to connect states with $|\Delta v| \leq k$, where $k$ is the expansion order. The additional dipole moment derivatives can be fit to data or calculated by \emph{ab initio} methods.

The internal state dynamics under the influence of blackbody excitation and radiative decay are determined by a set of rate equations for the population, $n_i$, of rovibrational state $i$:
\begin{equation}
    \frac{dn_i}{dt} = - \sum_j R_{ij}n_i - \sum_{j<i} A_{ij}n_i + \sum_j R_{ji}n_j + \sum_{j>i} A_{ji}n_j
    \label{eqn:rateeqn},
\end{equation}
where $\sum_{j<i}$ implies a sum over all states lower in energy than state $i$.

Fig. \ref{fig:rateeqnplots} compares the solution to eqn. \ref{eqn:rateeqn} for CaOH in a 300K environment including both bending and stretching vibrations (see Fig. \ref{fig:leveldiagram}), with a diatomic analogue which has the same stretching vibrational constants but no bending vibration. For comparison, we also plot the equilibrium vibrational populations given by
\begin{equation}
    P(\{v\}_i) = \frac{1}{Z} d_i \exp \left(\frac{\sum_n v_{ni} \hbar \omega_n}{k_B T} \right),
\end{equation}
where
\begin{equation}
    Z = \sum_i d_i \exp \left(\frac{\sum_n v_{ni} \hbar \omega_n}{k_B T} \right)
\end{equation}
is the partition function, $i$ sums over all vibrational states $\{v\}_i$, $d_i$ is the degeneracy of the $i$th vibrational state, $n$ sums over the vibrational modes, and $\omega_n$ is the frequency of the $n$th mode. As shown in Fig. \ref{fig:rateeqnplots}, after an initial equilibration time the vibrational populations (including population in all rotational levels of a vibrational state) converge to these equilibrium values. The individual rotational levels (dashed curves) take much longer to equilibrate.

Note that even after the populations reach steady state, individual molecules will continue to undergo transitions between rovibrational states. At any time, individual state lifetimes are given by the rate at which population leaves the state due to blackbody excitation and radiative decay. In terms of the rates in eqn. \ref{eqn:rateeqn}, the lifetime of a single rovibrational state $i$ is $\tau_i = (\sum_{j}R_{ij}+\sum_{j<i}A_{ij})^{-1}$.

\subsection{Other experimental limitations to lifetime}

In addition to blackbody thermalization and radiative decay, there are other sources of population loss in experiments with trapped molecules. One possible mechanism is trap loss due to collisions with background gas in the vacuum chamber, which generally leads to exponential population decay on typical timescales of 1 to 100~s in experiments with optically or magnetically trapped molecules. In addition, collisions with other trapped molecules can give density-dependent losses \cite{cheuk_2020, christianen2019photoinduced, gregory2020loss, Liu2020photo}, though these are not considered in this work due to the relatively low molecule number densities achieved in the experimental data described below.

In another mechanism, photon scattering from optical traps can change the internal state of optically trapped molecules in a process known as Raman scattering. While a single Raman scattering event manifests as loss from the rotational state of interest, Raman scattering may be suppressed by a large margin compared to Rayleigh scattering, which preserves the internal state \cite{cline1994laser}. For linearly polarized trapping light, Rayleigh scattering arises from state-preserving terms proportional to the square of the scalar polarizability of the molecule, $\alpha_0$, while Raman scattering arises from state-changing terms related to the tensor polarizability, $\alpha_2$. Therefore, the ratio of Raman to Rayleigh scattering is on the order of $(\alpha_2/\alpha_0)^2$, which is $\sim 1\%$ for CaOH. Because the Rayleigh scattering rate is $\sim$1 Hz for the optically trapped CaOH as described below, Raman scattering occurs on timescales of $\sim 30$ s and is neglected from the model.

\section{Measurement of C\lowercase{a}OH Lifetimes}
\label{sec:caoh}

In this section we describe a measurement of the blackbody lifetime of the $\widetilde{X}^2\Sigma^+(000)$ vibrational ground state of CaOH, as well as blackbody and radiative lifetimes of the $\widetilde{X}^2\Sigma^+(01^10)$ and $\widetilde{X}^2\Sigma^+(100)$ states. This is achieved by fitting experimental state lifetime data to a rate equation model of the full experimental sequence.

\subsection{Experimental protocol}

\begin{figure}
    \centering
    \includegraphics{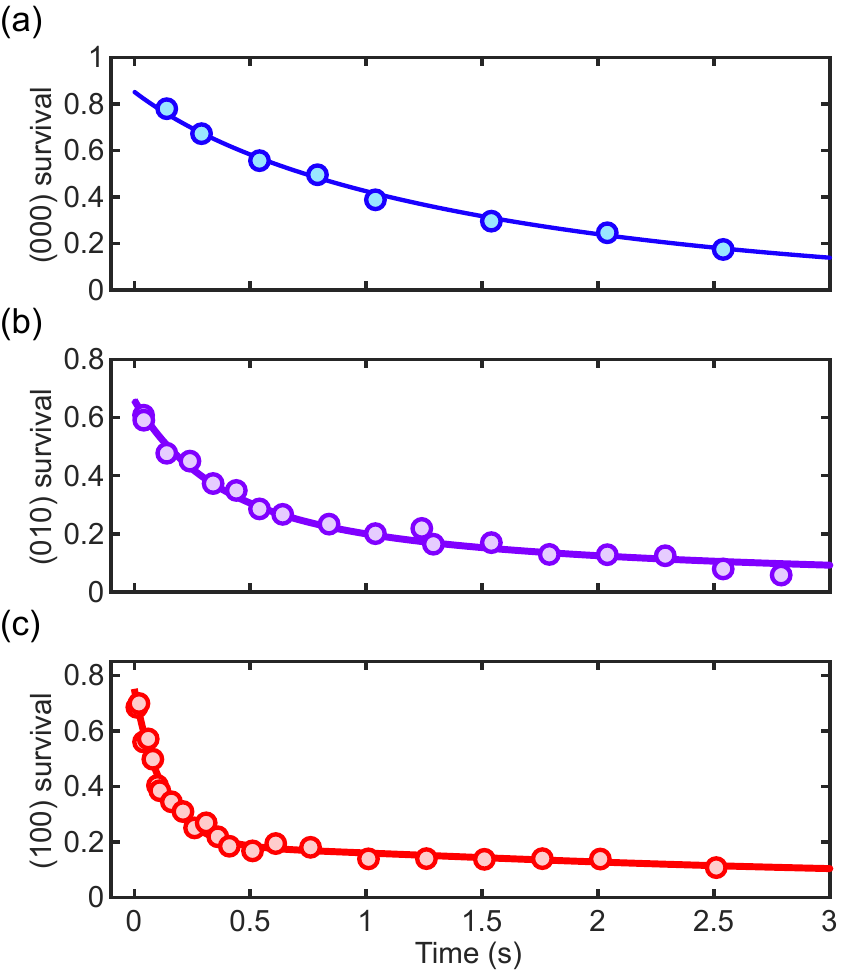}
    \caption{Lifetime measurements of the (a) $\widetilde{X}^2\Sigma^+(000)$, (b) $\widetilde{X}^2\Sigma^+(000)$, and (c) $\widetilde{X}^2\Sigma^+(000)$ states in optically trapped CaOH. Points are experimental data reported in Ref. \cite{hallas2022optical}, and curves are fits to the rate equation model described in Sec. \ref{sec:caoh}.}
    \label{fig:viblifetimes}
\end{figure}

The experimental protocol for measuring CaOH vibrational state lifetimes is described in detail in Ref. \cite{hallas2022optical}. In brief, CaOH molecules are loaded from a magneto-optical trap (MOT) \cite{vilas2022magneto} into an optical dipole trap (ODT) with a trap depth of $\sim$600~$\mu$K. The molecules are then prepared in the $N=1, p=-1$ level of either the $(000)$, $(01^10)$, or $(100)$ vibrational state, and held in the ODT for a variable time. The population remaining in detectable states (listed in Tab. \ref{tab:detectablestates}) is then measured. The population as a function of time for these three vibrational states is shown in Fig. \ref{fig:viblifetimes}, along with fits to the rate equation model, as described below. The radiative, blackbody, and vacuum lifetimes for these states are determined from the rate equation fit.

\begin{table}[]
    \centering
    \begin{tabular}{c|l|l|c|c|l|l}
    \hline
    \hline
    $(v_1 v_2^\ell v_3)$ & $N$ & \multicolumn{1}{c|}{$J$} & & $(v_1 v_2^\ell v_3)$ & $N$ & \multicolumn{1}{c}{$J$}  \\
    \hline
    (000) & 1 & $1/2,3/2$ & & $(02^00)$ & 1 & $1/2,3/2$ \\
    (100) & 1 & $1/2,3/2$ & & $(02^20)$ & 2 & $3/2$ \\
    (200) & 1 & $1/2,3/2$ & & $(11^10)$ & 1 & $1/2,3/2$ \\
    (300) & 1 & $1/2,3/2$ & & $(11^10)$ & 2 & $3/2$ \\
    $(01^10)$ & 1 & $1/2,3/2$ & & $(12^00)$ & 1 & $1/2,3/2$ \\
    $(01^10)$ & 2 & $3/2$ & & $(12^20)$ & 2 & $3/2$\\
    \hline
    \hline
    \end{tabular}
    \caption{Rovibrational states detected while imaging the trap after the lifetime measurements. In all cases only negative parity ($p=-1$) states are detected.}
    \label{tab:detectablestates}
\end{table}

\subsection{Rate equations}

The internal state dynamics of CaOH molecules in the experiment are described by the rate equations in eqn. \ref{eqn:rateeqn}, plus a term accounting for vacuum loss:
\begin{equation}
    \frac{dn_i}{dt} = - \sum_j R_{ij}n_i - \sum_{j<i} A_{ij}n_i + \sum_j R_{ji}n_j + \sum_{j>i} A_{ji}n_j - \frac{n_i}{\tau_\text{vac}}
    \label{eqn:rateeqn2},
\end{equation}
where $\tau_\text{vac}$ is the (state-independent) vacuum lifetime. In the experiment, the in-vacuum coils used to form the MOT are cooled to $\sim$0$^\circ$C to minimize outgassing and improve the vacuum pressure. The solid angle of the coils is $0.57\times 4\pi$ sr, so the blackbody rates $R_{ij}$ (eqn. \ref{eqn:BBR}) include two terms such that 57\% of the blackbody radiation seen by the molecules is at 273~K while the remaining 43\% is at room temperature (295~K). To preserve normalization of the population vector when integrating the rate equations, a ``lost'' population is also included according to
\begin{equation}
    \frac{dn_\text{loss}}{dt} = \sum_i \frac{n_i}{\tau_\text{vac}}.
    \label{eqn:nloss}
\end{equation}
Other sources of loss discussed in Sec. \ref{sec:RateEquations}B are expected to be negligible in this system.

While hyperfine structure is spectroscopically unresolved in the experiment and can therefore be omitted from the model, the photon-cycling detection does distinguish between the spin-rotation ($J$) sublevels in certain rotational manifolds. We therefore use a basis of rovibronic states described by the quantum numbers $\lvert v_1, v_2, \ell, N, J, p\rangle$, where $v_1$ and $v_2$ are the vibrational quantum numbers of the Ca--O stretch and Ca--O--H bending modes, $\ell$ is the magnitude of the vibrational angular momentum projected onto the molecular axis, $N$ is the rotational quantum number, $J$ is the total electronic angular momentum, and $p$ is the parity. We neglect the O--H stretching mode because its frequency is far from the peak of the blackbody spectrum at room temperature.  All states with $\ell=0$ have $\Sigma^+$ symmetry and their parity is $p=(-1)^N$. For states with $\ell \neq 0$, each rotational manifold has two opposite parity levels, $\lvert \ell, N, \pm\rangle = 2^{-1/2}\left\{\lvert \ell, N\rangle \pm (-1)^{N-\ell}\lvert -\ell, N\rangle\right\}$. Because $J$ is included in the basis, the H\"onl-London factors in eqn. \ref{eqn:hlvib} are modified to read:
\begin{align}
    S^\text{rot}_{ij} &= \delta_{p_i,-p_j}(1+\delta_{\ell_i 0} + \delta_{\ell_j 0} - 2 \delta_{\ell_i0}\delta_{\ell_j0}) \nonumber \\ 
    &\times (2N_i+1)(2N_j+1)
    \begin{pmatrix}
    N_i & 1 & N_j \\
    -\ell_i & \ell_i-\ell_j & \ell_j
    \end{pmatrix}
    ^2 \nonumber \\
    &\times (2J_i+1)(2J_j+1)
    \begin{Bmatrix}
    N_j & J_j & S \\
    J_i & N_i & 1
    \end{Bmatrix}
    ^2,
\label{eqn:hlvibJ}
\end{align}

The rate equations are solved by numerically integrating eqns. \ref{eqn:rateeqn2}-\ref{eqn:nloss} (sometimes with additional terms described below) applied to a population vector including all states with $v_1 \leq 2$, $v_2 \leq 2$, $N\leq 5$. The populations of the highest states included in each rotational and vibrational ladder are typically $<$5\% over the experimental timescale ($t\lesssim5$ s), implying that inclusion of higher-energy states is unnecessary.

Note that for the CaOH modeling we employ the double-harmonic approximation described above and omit anharmonic terms in the potential. This limits the number of required fit parameters and makes the fitting process tractable. Because the majority of the dynamics considered in this work occur in vibrational levels near the bottom of the molecular potential, the harmonic approximation is expected to be reasonably appropriate.  Nonetheless, the fitted values of $|d\vec{\mu}_e/dQ_i|$ should be interpreted as ``effective'' parameters that include the contribution of anharmonic effects on the measured $(000)$, $(010)$, and $(100)$ lifetimes. These contributions are expected to be relatively small: the \textit{ab initio} calculations performed in Sec. \ref{sec:abinitio} indicate that anharmonicity in the molecular potential is expected to have a $\sim$10-20\% effect on the vibrational lifetimes.

\subsection{Modeling the experimental sequence}

To accurately determine the lifetimes of the $\widetilde{X}(000)$, $(01^10)$, and $(100)$ states of CaOH, we use rate equations to model the full experimental sequence used to perform the lifetime measurements. This approach accounts for blackbody dynamics that occur prior to the lifetime measurement hold time, ensuring accurate initial conditions. It is found that significant population accumulates in dark vibrational levels during the ODT loading, imaging, and optical pumping steps, justifying their inclusion in the model.

\begin{figure}
    \centering
    \includegraphics[width= 1\columnwidth]{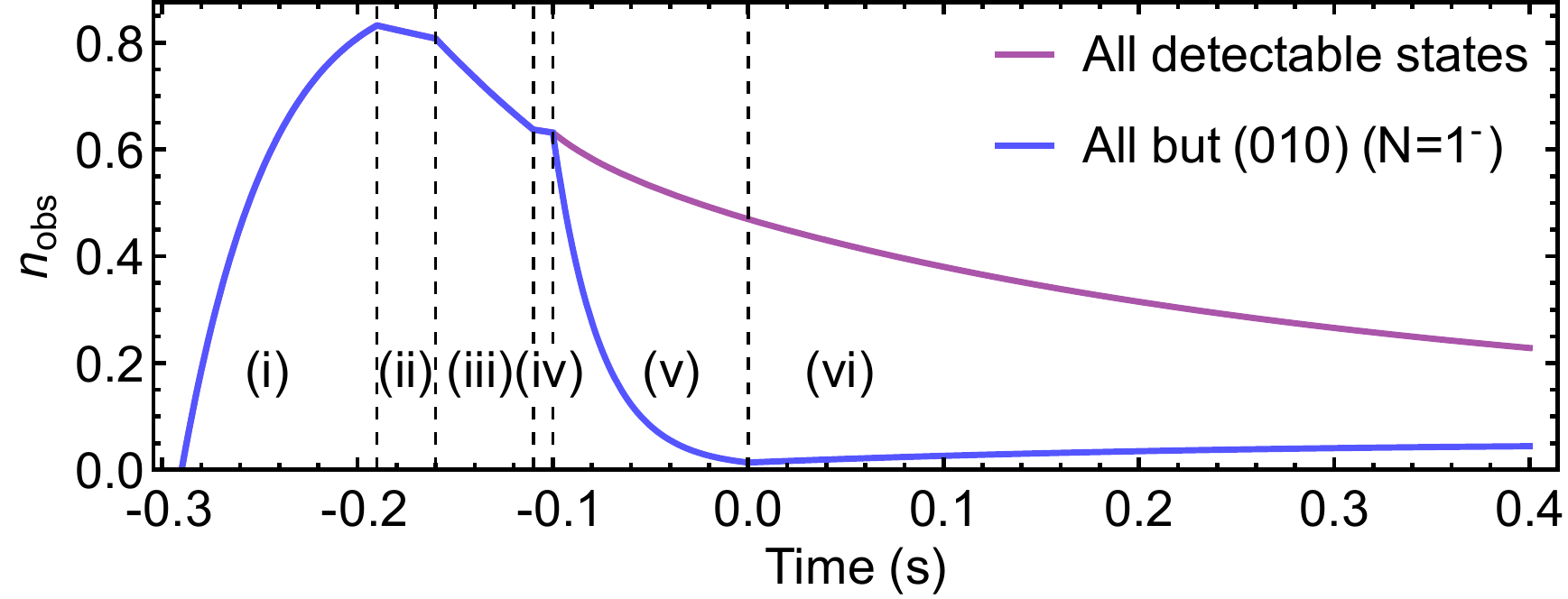}
    \caption{Population in detectable states (purple curve) and detectable states except $\widetilde{X}(010)(N=1^-)$ (blue curve) calculated from the best-fit rate equation for the $\widetilde{X}(010)$ lifetime sequence. The steps in the sequence are (i) ODT loading (ii) hold (iii) first image (iv) hold (v) optical pumping into $\widetilde{X}(010)$ and (vi) lifetime hold. $t=0$ here is the same as in Fig. \ref{fig:viblifetimes}.}
    \label{fig:sequence}
\end{figure}

The sequence is modeled as follows, and is chosen to exactly match the experimental protocol unless otherwise specified. As an example, Fig. \ref{fig:sequence} shows the detectable population throughout the $\widetilde{X}(01^10)$ lifetime measurement sequence, illustrating the key steps in the experimental protocol and the corresponding rate equation model. The simulation is initialized with all molecules in the untrapped state, $n_\text{loss}(t=0) = 1$. This population is pumped into the trapped state $\widetilde{X}^2\Sigma^+(000)(N=1, J=3/2)$ with a characteristic rate of $R_\text{load}=1/(48.5 ~\text{ms})$, fit from the experimental data (Fig. 2 of Ref. \cite{hallas2022optical}). This optical pumping step, as well as the trap imaging described later, is performed with optical cycling light in a single-frequency (SF) cooling configuration, which simultaneously cools the molecules inside the optical trap \cite{hallas2022optical}. It is assumed that population accumulates entirely in the $\widetilde{X}^2\Sigma^+(000)(N=1, J=3/2)$ level during SF cooling since this is the furthest-detuned ground-state level. Additionally, coherent dark states are formed within the $J=3/2$ manifold during SF cooling \cite{caldwell2019deep}. Therefore, whenever SF cooling light is on, an additional term is added to the rate equations which pumps population from $\widetilde{X}^2\Sigma^+(000)(N=1, J=1/2)$ to $\widetilde{X}^2\Sigma^+(000)(N=1, J=3/2)$ at a rate corresponding to the SF scattering rate of $R_\text{sc} = 45\times 10^3$~s$^{-1}$ \cite{hallas2022optical,tarbutt2015magneto}.

As soon as the molecules are trapped, their internal states begin evolving according to eqns. \ref{eqn:rateeqn2}-\ref{eqn:nloss}. Prior to ODT loading, these dynamics are suppressed because any molecule which is blackbody-excited out of the cycling scheme will not be cooled/loaded into the ODT (this is enforced in the rate equations because the population $n_\text{loss}$ is not subject to blackbody terms). After the ODT loading time of 100~ms, the loading is turned off and the remaining molecules evolve through each step in the experimental sequence. First, the molecules are held in the ODT and propagate according to eqns. \ref{eqn:rateeqn2}-\ref{eqn:nloss} for 30 ms. Next, the molecules undergo 50 ms of ``normalization'' SF imaging, used to determine the number of molecules loaded into the trap. In addition to the dynamics in eqns. \ref{eqn:rateeqn2}-\ref{eqn:nloss}, during this time molecules in detectable states are pumped into dark vibrational levels $n_\text{loss}$ (which are high-lying and therefore neglected for the remainder of the simulation) at a rate corresponding to a branching ratio of $v_\text{dark} = 8.5\times10^{-5}$ and a scattering rate of $R_\text{sc} = 45 \times 10^3$ s$^{-1}$ \cite{vilas2022magneto,hallas2022optical}. Additionally, molecules in repumped vibrational levels are pumped back into $\widetilde{X}^2\Sigma^+(000)(N=1)$ during the imaging at a rate of $10^3$~s$^{-1}$, and molecules in $\widetilde{X}^2\Sigma^+(000)(N=1, J=1/2)$ are pumped into $\widetilde{X}^2\Sigma^+(000)(N=1, J=3/2)$ as described above.

After the normalization image, the rate equation propagation proceeds differently for the $\widetilde{X}(000)$, $\widetilde{X}(010)$, and $\widetilde{X}(100)$ states. For the $\widetilde{X}(000)$ data, the populations propagate according to eqns. \ref{eqn:rateeqn2}-\ref{eqn:nloss} for a variable time. For comparison to the experimental data in Fig. \ref{fig:viblifetimes}, $t=0$ occurs immediately after the imaging light is turned off. The detectable population, $n_\text{obs}(t) = \sum_{i\in\{\text{det}\}} n_i(t)$, for the best fit parameters (see below) is plotted as a solid curve in Fig. \ref{fig:viblifetimes}(a). Note that here we make the approximation that the molecules are detected instantaneously, whereas in the experiment the final image is collected over 50~ms. We have run the model including finite imaging time and confirmed that this approximation does not change the results.

For the $\widetilde{X}(010)$ and $\widetilde{X}(100)$ lifetimes the molecules propagate according to eqns. \ref{eqn:rateeqn2}-\ref{eqn:nloss} for a short time (10 ms for $(010)$ and 90 ms for $(100)$) following the normalization image. This is followed by optical pumping into the desired vibrational state, which is modeled by adding a term that pumps population in detectable states into the excited vibrational level at a rate $R_{\text{pump},i}=R_\text{sc}v_iS_i$, where $R_\text{sc} = 45 \times 10^3$ s$^{-1}$ is the SF scattering rate, $v_{(010)} = 8.2 \times 10^{-4}$ is the vibrational branching ratio (VBR) to $\widetilde{X}(010)(N=1^-)$, and $v_{(100)} = 4.75 \times 10^{-2}$ is the VBR to $\widetilde{X}(100)(N=1)$ \cite{hallas2022optical}. The rotational branching factors for $\widetilde{X}(010)$ are $S_{1/2}=0.73$ and $S_{3/2}=0.27$ to $J=1/2$ and $J=3/2$, respectively, and for $\widetilde{X}(100)$ they are $S_{1/2}=2/3$ and $S_{3/2}=1/3$ \cite{baum2021establishing}. Loss to dark vibrational states at a rate $R_\text{dark} = R_\text{scatt}v_\text{dark}$, where $v_\text{dark} = 8.5 \times 10^{-5}$, is also included. The $\widetilde{X}(010)$ pumping is turned on for 100 ms, and the $\widetilde{X}(100)$ pumping for 2 ms.

After optical pumping, the populations propagate according to eqns. \ref{eqn:rateeqn2}-\ref{eqn:nloss} for a variable time. In Fig. \ref{fig:viblifetimes}, $t=0$ is defined as immediately after the optical pumping/state transfer light is turned off. The observable population, as well as the detectable population in all states besides $\widetilde{X}(010)(N=1^-)$, is plotted in Fig. \ref{fig:sequence} for the full $\widetilde{X}(010)$ lifetime sequence.

\subsection{Fitting and results}

We fit the model results to the experimental data as follows. The rate equations for the $\widetilde{X}(000)$, $\widetilde{X}(010)$, and $\widetilde{X}(100)$ measurement sequences are first propagated for given values of the vacuum lifetime, $\tau_\text{vac}$, and the dipole derivatives $|d\vec{\mu}_e/dQ_1|$ (symmetric stretch) and $|d\vec{\mu}_e/dQ_2|$ (bend). We then calculate the observable population, $n_\text{obs}(t) = \sum_{i\in\{\text{det}\}} n_i(t)$, where $\{\text{det}\}$ is the subset of states which are detectable, i.e., the $(N=1, J=1/2^-)$, $(N=1, J=3/2^-)$, and $(N=2, J=3/2^-)$ levels of repumped vibrational states (Tab. \ref{tab:detectablestates}). Finally, the results are scaled by constant prefactors $a_{\{v\}_i}$, and a constant offset $a_{\text{off},\{v\}_i}$ (present due to imperfections in the imaging background subtraction, or imperfect state preparation) is added to each of the three traces. The resulting fit functions depend on a total of 9 fit parameters:
\begin{align}
    f_{(000)}(t) &= a_{(000)}n_{\text{obs},(000)}\left(t,    \left|\frac{d\vec{\mu}_e}{dQ_1}\right|,
    \left|\frac{d\vec{\mu}_e}{dQ_2}\right|,
    \tau_\text{vac}\right) + a_{\text{off},(000)}, \\
    f_{(010)}(t) &= a_{(010)}n_{\text{obs},(010)}\left(t,    \left|\frac{d\vec{\mu}_e}{dQ_1}\right|,
    \left|\frac{d\vec{\mu}_e}{dQ_2}\right|,
    \tau_\text{vac}\right) + a_{\text{off},(010)}, \\
    f_{(100)}(t) &= a_{(100)}n_{\text{obs},(100)}\left(t,
    \left|\frac{d\vec{\mu}_e}{dQ_1}\right|,
    \left|\frac{d\vec{\mu}_e}{dQ_2}\right|,
    \tau_\text{vac}\right) + a_{\text{off},(100)}.
\end{align}
These functions are plotted as solid curves in Fig. \ref{fig:viblifetimes} for the best fit parameters determined below.

To perform the fit, the fit parameters are scanned on a discrete 9-dimensional grid and the fit functions are calculated at each point. Step sizes of 0.025~D for the dipole derivatives, 0.5~s for the vacuum lifetime, 0.05 for the amplitudes, and 0.02 for the offset are used. The sum of squared errors,
\begin{align}
S & = \sum_i \left[f^{\text{exp}}_{(000)}(t_i) - f_{(000)}(t_i)\right]^2 + \sum_i \left[f^\text{exp}_{(010)}(t_i) - f_{(010)}(t_i)\right]^2 \nonumber \\
& + \sum_i \left[f^\text{exp}_{(100)}(t_i) - f_{(100)}(t_i)\right]^2,
\end{align}
where $f_{\{v\}}^\text{exp}(t_i)$ is the experimentally measured survival at time $t_i$ and $i$ sums over the experimental data points, is then calculated for each point on the 9-dimensional grid. The optimal fit parameters are determined by fitting the 10,000 lowest $S$ values (approximately 1\% of the full grid) to a second-order polynomial. The fit is constrained to this subset of points in order to minimize the effect of higher-order curvature of the error surface. The parameter errors determined from this fit are negligible compared to error sources described below.

\begin{figure*}
    \centering
    \includegraphics[width=0.8\textwidth]{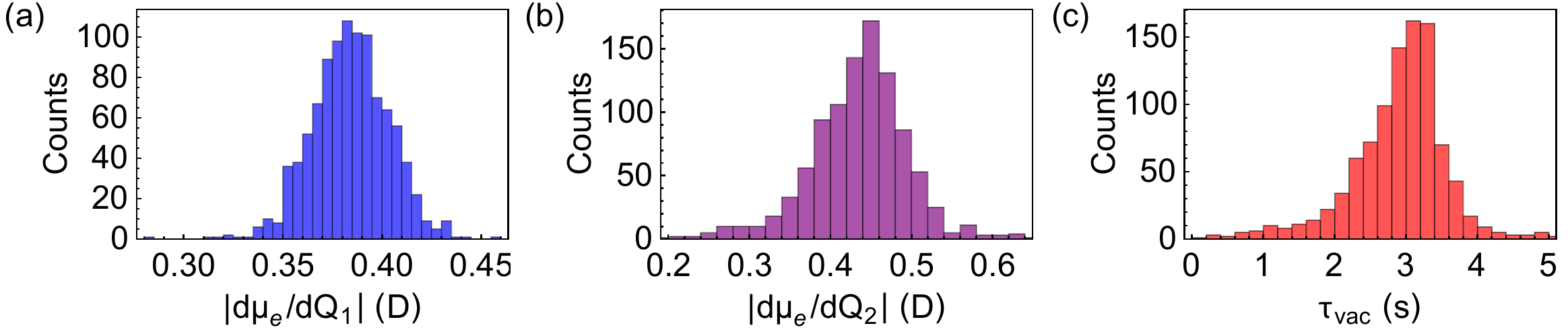}
    \caption{Histograms of the parameters (a) $|d\vec{\mu}_e/dQ_1|$, (b) $|d\vec{\mu}_e/dQ_2|$, and (c) $\tau_\text{vac}$ generated by fitting the rate equation model to $\sim$1000 synthetic data sets accounting for experimental error bars, as described in the text.}
    \label{fig:histograms}
\end{figure*}

The primary source of uncertainty in the fitted lifetimes is due to correlations between parameters, for example between the bending mode lifetime and fit amplitude, $|d\vec{\mu}_e/dQ_2|$ and $a_{(010)}$. These correlations have the potential to make the fit results overly sensitive to the precise values of the individual data points. Additionally, the $\widetilde{X}(010)$ radiative decay occurs on a similar timescale to blackbody excitation and vacuum loss, making it difficult to isolate. To account for these factors and adequately estimate the parameter errors, we repeat the fitting procedure $\sim$1000 times, each time fitting to a ``synthetic'' data set produced by sampling the value of each point, $f_{\{v\}}^\text{exp}(t_i)$, from a normal distribution corresponding to the mean and standard deviation of the measured data point (i.e., of the data plotted in Fig. \ref{fig:viblifetimes}). This is meant to approximate the variation in the fit parameters expected if the experiment were to be repeated 1000 times. Histograms of the resulting fit parameters are shown in Fig. \ref{fig:histograms}. The parameters and error bars are determined by taking the median of the histogram as the center value and the middle 68\% of the distribution as the $1\sigma$ confidence interval. Results for each of the fit parameters are shown in Tab. \ref{tab:rateresults}.

\begin{table}
    \centering
    \begin{tabular}{ c | c | c }
    \hline
    \hline
    Parameter & Median & 68\% Conf. Int.  \\
    \hline
    $|d\vec{\mu}_e/dQ_1|$ & 0.384 D & (0.365, 0.404) D \\
    $|d\vec{\mu}_e/dQ_2|$ & 0.440 D & (0.380, 0.486) D \\
    $\tau_\text{vac}$ & 2.99 s & (2.33, 3.43) s \\
    $a_{(000)}$ & 1.36 & (1.29, 1.43) \\
    $a_{(010)}$ & 1.31 & (1.18, 1.40) \\
    $a_{(100)}$ & 1.18 & (1.11, 1.24) \\
    $a_{\text{off},(000)}$ & -0.015 & (-0.044, 0.018) \\
    $a_{\text{off},(010)}$ & 0.045 & (0.032, 0.061) \\
    $a_{\text{off},(100)}$ & 0.063 & (0.049, 0.076) \\
    \hline
    \hline
    \end{tabular}
    \caption{Fit parameters and 68\% confidence intervals for the rate equation fit.}
    \label{tab:rateresults}
\end{table}

\begin{table}
    \centering
    \bgroup
    \def\arraystretch{2}
    \begin{tabular}{ c | c c c}
    \hline
    \hline
    State & $\tau_\text{spont}$ & $\tau_\text{bbr}$ & $\tau_\text{tot}$  \\
    \hline
    $\widetilde{X}(000)$ & -- & $1.3^{+0.3}_{-0.2}$ & $0.90^{+0.20}_{-0.16}$ \\
    $\widetilde{X}(010)$ & $0.72^{+0.25}_{-0.13}$ & $0.95^{+0.26}_{-0.16}$ & $0.36^{+0.11}_{-0.07}$ \\
    $\widetilde{X}(100)$ & $0.19^{+0.03}_{-0.03}$ & $0.81^{+0.09}_{-0.08}$ & $0.14^{+0.02}_{-0.02}$\\
    \hline
    \hline
    \end{tabular}
    \egroup
    \caption{State lifetimes and 68\% confidence intervals determined from the rate equation fit. The vacuum lifetime is constant for all states and fits to $\tau_\text{vac}=3.0^{+0.4}_{-0.7}$~s.}
    \label{tab:lifetimeresults}
\end{table}

The state lifetimes are calculated from the fitted parameters and eqns. \ref{eqn:spont}-\ref{eqn:BBR} by summing over all allowed transitions, i.e.,
\begin{align}
    \tau_{\text{spont},i} &= \left(\sum_{j<i} A_{ij}\right)^{-1}, \\
    \tau_{\text{bbr},i} &= \left(\sum_j R_{ij}\right)^{-1},  \\
    \tau_{\text{tot},i} &= \left(\frac{1}{\tau_{\text{spont},i}}+\frac{1}{\tau_{\text{bbr},i}}+\frac{1}{\tau_{\text{vac}}}\right)^{-1}.
\end{align}
The results for CaOH are given in Tab. \ref{tab:lifetimeresults} and agree well with \emph{ab initio} calculations described in Sec. \ref{sec:abinitio}, below.

The amplitudes $a_{(000)}$, $a_{(010)}$, and $a_{(100)}$ can be understood as scale factors between the calculated populations, which are normalized to 1, and the measured populations, which are normalized to the number of molecules detected in the first image. From the fitted model, the detectable population at the start of the first image is $n_\text{obs}(t= t_\text{img}) = 0.81$, so the model results need to be scaled by $1/0.81 = 1.24$ to match the experimental data. The fitted amplitudes in Tab. \ref{tab:rateresults} are in good agreement with this expectation.

\section{Ab initio calculations for alkaline-earth monohydroxides}
\label{sec:abinitio}

In this section, we present calculations of radiative and blackbody lifetimes of the $\widetilde{X}(000)$, $\widetilde{X}(010)$ and $\widetilde{X}(100)$ states of CaOH, as well as lifetimes of the $\widetilde{X}(010)$ states of SrOH and YbOH. The $\widetilde{X}(010)$ bending mode in each of these molecules is expected to have useful applications \cite{augustovivcova2019ultracold, kozyryev2017precision, kozyryev2021enhanced, anderegg2023quantum}, with the radiative lifetime setting a limit on the achievable interrogation times for such experiments.

\subsection{Computational details and results}

The calculation of a spontaneous decay rate (in a.u.)
\begin{equation}
    \Gamma_\text{sp} = \frac{4\omega^3  |\vec{\mu}_v|^2}{3c^3},
    \label{eqn:gammasp}
\end{equation}
involves the transition dipole moment $|\vec{\mu}_v|$ and the energy difference $\omega$ between two vibrational states.
Here $c$ is the speed of light. Note that atomic units (a.u.) are used for the remainder of this work.
The present discrete variable representation (DVR) \cite{Colbert1992, zhang2021accurate} calculations expand the vibrational wave functions in terms of real space basis functions on a grid with 21 evenly spaced points in the range $[-4.0Q,4.0Q]$ for the bending modes and the metal-oxygen stretching mode as well as 28 points in the range $[-6.8Q,4.0Q]$ for the O-H stretching mode, in which $Q$ represents the corresponding dimensionless normal mode of the $\widetilde{X}^2\Sigma^+$ state.
The transition dipole moment between $\widetilde{X}^2\Sigma^+(010)$ and $\widetilde{X}^2\Sigma^+(000)$ (for example) was calculated as 
\begin{equation}
    \label{expectation}
    \vec{\mu}_v = \langle \chi_{000}(\mathbf{R}) | \vec{\mu} (\mathbf{R}) | \chi_{010} (\mathbf{R})\rangle ,
\end{equation}
in which $\mathbf{R}$ represents the normal coordinates, $\chi_{000}(\mathbf{R})$ and $\chi_{010}(\mathbf{R})$ are the vibrational wave functions, and $\vec{\mu} (\mathbf{R})$ is the dipole-moment function of the $\widetilde{X}^2\Sigma^+$ state. 
The dipole-moment functions were obtained by fitting calculated dipole moment values for the structures near the equilibrium structure into a polynomial of normal coordinates. We fitted 625 dipole-moment values computed on a grid, which consists of 5 evenly spaced points in the range $[-0.2Q,0.2Q]$ for each normal mode, into a fourth-order polynomial function in terms of the normal coordinates.  
The dipole-moment calculations on this grid were performed using the same computational methods as those used for calculations of potential energy surfaces in Ref. \cite{zhang2021accurate}, i.e., the equation-of-motion electron-attachment coupled-cluster singles and doubles (EOMEA-CCSD) method \cite{EOMCC1Stanton1993,EOMEA_Nooijen95} and the correlation-consistent quadruple-zeta (QZ) basis sets for CaOH and triple-zeta (TZ) basis sets for SrOH and YbOH \cite{dunning1989gaussian,koput2002ab,de2001parallel,hill2017gaussian,lu2016correlation}.
Detailed information about the basis sets, the frozen orbitals, and the potential energy surfaces for the DVR calculations have been documented in Ref. \cite{zhang2021accurate}.

\begin{table}
  \begin{center}
  \begin{tabular}{ccccccc}
    \hline \hline
     &~&  $|\vec{\mu}_v|$   &~& $\tau_\text{sp}$ &~& $\tau$ (300 K)\\
    \hline
    CaOH  &~& 0.284  &~& 876 &~& 409 \\
    SrOH  &~& 0.268  &~& 902 &~& 439 \\
    YbOH  &~& 0.305  &~& 1020 &~& 440 \\
    \hline \hline
  \end{tabular}
    \caption{The $\widetilde{X}^2\Sigma^+(000)-\widetilde{X}^2\Sigma^+(010)$ transition dipole moments (in debye), spontaneous lifetimes (in ms), and overall lifetimes at 300 K (in ms) for the $\widetilde{X}^2\Sigma^+(010)$ states of CaOH, SrOH, and YbOH.}
    \label{tab:casryb}
  \end{center}
\end{table}

\begin{table}
  \begin{center}
  \begin{tabular}{ccccccc}
    \hline \hline
      &~&  $|\vec{\mu}_v|$   &~& $\tau_\text{sp}$ &~& $\tau$ (300 K)\\
    \hline
    $\widetilde{X}^2\Sigma^+(100)$  &~& 0.295  &~& 161  &~& 141 \\
    $\widetilde{X}^2\Sigma^+(000)$  &~& -      &~& -    &~& 1143 \\
    \hline \hline
  \end{tabular}
    \caption{Calculated lifetimes of the $\widetilde{X}^2\Sigma^+(100)$ and $\widetilde{X}^2\Sigma^+(000)$ states of CaOH. $|\vec{\mu}_v|$ is the $\widetilde{X}^2\Sigma^+(000)-\widetilde{X}^2\Sigma^+(100)$ transition dipole moment (in debye), $\tau_\text{sp}$ is the spontaneous lifetimes (in ms), and $\tau$~(300~K) is the overall lifetime at 300~K.}
    \label{tab:caoh}
  \end{center}
\end{table}

\begin{table*}
  \begin{center}
  \begin{tabular}{ccccccccccccc}
    \hline \hline
    Vibrational & \multicolumn{2}{c}{$|\vec{\mu}_v|$} &~& \multicolumn{2}{c}{$|d\vec{\mu}_e/dQ|$}  &~& \multicolumn{3}{c}{$\tau_\text{sp}$} &~& \multicolumn{2}{c}{$\tau$ (300 K)}\\
        state              & Calc. (H) & Calc. &~& Calc. & Exp. &~& Calc. (H) & Calc. & Exp. &~& Calc. & Exp.\\
    \hline
    (010) & 0.301 & 0.284 &~& 0.426 & 0.440 &~& 0.78 & 0.88 & 0.72 &~& 0.41  & 0.41 \\
    (100) & 0.262 & 0.295 &~& 0.370 & 0.384    &~& 0.20 & 0.16 & 0.19 &~& 0.14  & 0.15 \\
    (000) & -     & -     &~& -    & -       &~& -   & -   & -   &~& 1.14 & 1.3 \\
    \hline \hline
  \end{tabular}
    \caption{Calculated transition dipole moments $|\vec{\mu}_v|$ (in debye), calculated and experimentally determined derivatives of the electronic dipole moment $d\vec{\mu}_e/dQ$ (in debye), spontaneous lifetimes (in s), and overall lifetimes at 300 K (in s)
    for the $\widetilde{X}^2\Sigma$ state of CaOH. ``(H)'' denotes results obtained using the harmonic approximation.  Experimental results quoted here ignore vacuum loss and are drawn from Tab. \ref{tab:rateresults} for the dipole derivatives and Tab. \ref{tab:lifetimeresults} for the lifetimes. The experimental values of $|d\vec{\mu}_e/dQ|$ are determined using the harmonic approximation as described in Sec. \ref{sec:caoh}.}
    \label{tab:harmonic_tab}
  \end{center}
\end{table*}

The overall lifetime was obtained by further including contributions from black-body radiation (BBR) induced transitions from the state of interest (e.g. $\widetilde{X}^2\Sigma^+(010)$) to higher excited states at 300 K using the formulae developed in Refs. \cite{vanhaecke2007precision,beterov2009quasiclassical}. The BBR decay rate of state $i$ can be evaluated by summing over the other states $i^{\prime}$,
\begin{equation}
    \Gamma_\text{BBR} = \sum_{i^{\prime}}\Gamma_\text{sp}(i\to i^{\prime})\frac{1}{e^{\omega_{ii^{\prime}}/k_B T}-1},
    \label{eqn:gammaBBR}
\end{equation}
where $k_B$ is the Boltzmann constant and $\omega_{ii^{\prime}}$ is the energy difference between $i$ and $i^{\prime}$ states.
In the present calculations, $i$ corresponds to the $\widetilde{X}^2\Sigma^+(010)$ state
and $\Gamma_\text{BBR}$ receives non-negligible contributions from the transitions to the $\widetilde{X}^2\Sigma^+(020)$ and $\widetilde{X}^2\Sigma^+(110)$ states. 
The calculated $\widetilde{X}^2\Sigma^+(000)-\widetilde{X}^2\Sigma^+(010)$ transition dipole moments ($|\vec{\mu}_v|$), spontaneous lifetimes ($\tau_\text{sp}$), and overall lifetimes ($\tau$) at 300~K for the $\widetilde{X}^2\Sigma^+(010)$ state of CaOH, SrOH, and YbOH are summarized in Tab. \ref{tab:casryb}.
The $\widetilde{X}^2\Sigma^+(010)$ states of these three molecules have long spontaneous lifetimes of around 1~s. 
The overall lifetimes at 300~K are around 400~ms. 
For the CaOH molecule, we also calculated the spontaneous and overall lifetimes for the $\widetilde{X}^2\Sigma^+(100)$ state and the overall lifetime for the ground $\widetilde{X}^2\Sigma^+(000)$ state. 
The results are summarized in Tab. \ref{tab:caoh}.
The lifetime for the $\widetilde{X}^2\Sigma^+(100)$ state is around 140~ms, considerably shorter than that of the $X^2\Sigma^+(010)$ state.
The computed lifetime of the vibrational ground state $\widetilde{X}^2\Sigma^+(000)$ amounts to around 1.14~s and is in good agreement with the measured value of 1.3~s. 

We also calculated the transition dipole moments and spontaneous lifetimes for the $\widetilde{X}^2\Sigma^+(010)$ and $\widetilde{X}^2\Sigma^+(100)$ states of CaOH using the harmonic approximation.
The transition dipole moment between the $v=1$ and $v=0$ states within the harmonic approximation, $\vec{\mu}_\text{H}$, can be evaluated as $\vec{\mu}_\text{H} = \frac{1}{\sqrt{2}} d\vec{\mu}_e/dQ$. The dipole derivative $d\vec{\mu}_e/dQ$ was obtained as the linear coefficients of the fitted dipole function.
A comparison between the DVR results (``Calc.''), the calculated results using the harmonic approximation [``Calc. (H)''], and the experimental measurements (``Exp.'') is given in Tab. \ref{tab:harmonic_tab}. 
The anharmonic contributions reduce the computed transition dipole moments for the $\widetilde{X}^2\Sigma^+(000)-\widetilde{X}^2\Sigma^+(010)$ transition
by around 6\% and hence increase the
computed lifetimes by around 13\%.
The computed spontaneous lifetimes in the harmonic approximation also agree well
with the measured ones and are within the uncertainty of the measured values.

\subsection{Benchmark analysis of the computational results} \label{benchmark}

To investigate the accuracy of the computed dipole-moment function, we calculated the dipole-moment function using the Hartree-Fock (HF), coupled-cluster singles and doubles (CCSD) \cite{Purvis82}, CCSD with a non-iterative triple [CCSD(T)] \cite{Raghavachari89}, and EOMEA-CCSD \cite{EOMEA_Nooijen95} methods with QZ and 5Z basis sets. 
The calculated vibrational transition dipole moments, electronic dipole derivatives, spontaneous lifetimes, and overall lifetimes for the $\widetilde{X}^2\Sigma^+(010)$ state of CaOH using these dipole moment functions are summarized in Tab. \ref{tab:dipole1}.
The EOM-CCSD/QZ results agree very well with the CCSD/QZ and CCSD(T)/QZ values. For example, the EOM-CCSD/QZ value for the transition dipole moment amounts to 0.284 debye, which is in close agreement with the CCSD/QZ value of 0.290 debye and the CCSD(T)/QZ value of 0.286 debye. The remaining electron-correlation contributions are expected to be small. 
The EOM-CCSD/QZ and EOM-CCSD/5Z results also agree with each other closely, indicating that
the remaining basis-set effects are small. 

\begin{table}
  \begin{center}
  \begin{tabular}{ccccccccc}
    \hline \hline
    Computational method    &~&  $|\vec{\mu}_v|$ &~&  $|d\vec{\mu}_e/dQ|$  &~& $\tau_\text{sp}$ &~& $\tau$ (300 K)\\
    \hline
    EOM-CCSD/QZ &~& 0.284 &~& 0.426 &~& 876 &~& 409 \\
    EOM-CCSD/5Z &~& 0.285 &~& 0.426 &~& 867 &~& 402 \\
    HF/QZ       &~& 0.294 &~& 0.437 &~& 813 &~& 372 \\
    CCSD/QZ     &~& 0.290 &~& 0.432 &~& 836 &~& 391 \\
    CCSD(T)/QZ  &~& 0.286 &~& 0.426 &~& 863 &~& 405 \\
    \hline \hline
  \end{tabular}
    \caption{Transition dipole moment (in debye), dipole derivative (in debye), spontaneous lifetime (in ms), and overall lifetime (in ms) of the $\widetilde{X}^2(010)$ state of CaOH obtained from dipole surfaces calculated using different methods.}
    \label{tab:dipole1}
  \end{center}
\end{table}

\begin{table}
  \begin{center}
  \begin{tabular}{ccccccc}
    \hline \hline
    Fitting range/Fitting order    &~&  $|\vec{\mu}_v|$   &~& $\tau_\text{sp}$ &~& $\tau$ (300 K)\\
    \hline
    $[-0.2Q,0.2Q]$/4th &~& 0.284  &~& 876 &~& 409 \\
    $[-2.0Q,2.0Q]$/4th &~& 0.287  &~& 854 &~& 399 \\
    $[-2.0Q,2.0Q]$/6th &~& 0.287  &~& 856 &~& 398 \\
    \hline \hline
  \end{tabular}
    \caption{Transition dipole moments $|\vec{\mu}_v|$ (in debye), spontaneous lifetimes $\tau_\text{sp}$ (in ms), and overall lifetimes $\tau$ at 300 K (in ms) for the $\widetilde{X}^2\Sigma^+(010)$ state of CaOH obtained from dipole moment functions fitted using the original and the enlarged data sets as well as with increased order of polynomial in the fitting.}
    \label{fit}
  \end{center}
\end{table}

We have examined the sensitivity of the computed results with respect to the grid points used to fit the dipole-moment function.
By looking into the contributions to the expectation value in eqn. (\ref{expectation}), we found that the contributions are mainly from the DVR basis functions within the range of $[-2.0Q,2.0Q]$ for each normal mode.
To ensure an accurate representation of this range, we performed dipole-moment calculations of on a grid of 625 points consisting of 5 evenly spaced points in $[-2.0Q, 2.0Q]$ for each normal mode. These computed dipole-moment values were added to the original data set in the fitting. The computed transition dipole moments, spontaneous lifetimes, and overall lifetimes using thus fitted dipole moment functions are summarized in Tab. \ref{fit}. There is a 3\% decrease of the computed lifetimes when using the dipole-moment function fitted using the enlarged data set. Increasing the order of polynomial to the 6th-order does not change the computed results significantly.

We have also used the EOM-CCSDT/TZ potential energy surface in the DVR calculation. 
This calculation gives a transition dipole moment $|\vec{\mu}_{01}|$ of 0.286 debye, slightly larger than the value of 0.284 debye obtained from calculations using the EOM-CCSD/TZ surfaces. Therefore, the remaining correlation effects on the potential energy surfaces play a minor role. We note that the computed spontaneous lifetime is proportional to the cubic power of energy difference between the vibrational states and thus is quite sensitive to this parameter. Since our calculated value of 356 cm$^{-1}$ is in close agreement with the measured value of 353 cm$^{-1}$, the corresponding error in the lifetime calculation is also expected to be small. 

Based the sources of errors discussed above, we give an error estimate of around 10\% for the computed transition dipole moments. This corresponds to an uncertainly of around 20\% for the computed lifetimes.

\section{Blackbody lifetimes of larger polyatomic molecules}
\label{sec:larger}

The study of CaOH, SrOH, and YbOH presented here 
has emphasized radiative lifetimes of the $\widetilde{X}^2\Sigma^+(010)$ states, which have been proposed as sensitive probes for
fundamental physics beyond the standard model (BSM) \cite{kozyryev2017precision, kozyryev2021enhanced, anderegg2023quantum} due to their parity-doublet structure. This structure also makes them amenable to other quantum science applications that require alignment of the molecule with external electric fields \cite{wall2013simulating,wall2015realizing,yu2019scalable}. The same parity-doublet structure is generically present in the vibrational ground state of nonlinear polyatomic molecules, where radiative decay is no longer a limitation and blackbody excitation is the dominant loss mechanism.
In this section, we extend our calculations to study nonlinear polyatomic molecules, focusing on blackbody-excitation induced lifetimes of the vibrational ground state. 
 
There have been many recent proposals \cite{kozyryev2016proposal,o2019hypermetallic,augenbraun2020molecular, klos2020prospects, dickerson2021franck, dickerson2021optical} and demonstrations \cite{mitra2020direct, zhu2022functionalization, mitra2022pathway, augenbraun2022high} of the potential of larger and/or more complex polyatomic molecules for laser cooling. On the other hand, as discussed qualitatively in Sec. \ref{sec:RateEquations} and quantitatively below, the ground state of CaOH has a significantly reduced blackbody lifetime compared to its close diatomic counterpart, CaF. This is due to the increased number of vibrational modes with frequencies near the peak of the 300~K blackbody spectrum in CaOH compared to CaF.
As laser cooling is extended to even larger polyatomic molecules, the increasing number of vibrational modes admits the possibility of very short blackbody lifetimes, in the same way that the extension from diatomic to linear triatomic molecules decreased the blackbody lifetime of CaOH compared to CaF. For example, in CaOCH$_3$, which has $N=6$ atoms and $3N-6 = 12$ vibrational modes, if transitions to even half the vibrational modes could be driven by blackbody radiation with strengths similar to the stretching and bending modes of CaOH, the blackbody lifetime of the ground state would be $\sim$0.5~s.
It thus is of interest to calculate blackbody lifetimes for laser-coolable, nonlinear polyatomic molecules. 

In this section, we present calculations of blackbody lifetimes for a number of complex polyatomic molecules of interest for laser cooling.
Since coupled-cluster calculations using the harmonic approximation presented in Sec. \ref{benchmark} have been shown to provide reliable results
for the lifetimes of CaOH, we have adopted the same level of theory for the calculations of non-linear molecules. DVR calculations for these molecules are beyond our present computational resources. 
 Interestingly, it is found that for a variety of these molecules the blackbody lifetimes are no shorter than that of CaOH.

\subsection{Computational details}

All the lifetime calculations for the non-linear molecules have been performed using the harmonic approximation. 
We have carried out EOMEA-CCSD calculations for the equilibrium structures and harmonic vibrational frequencies
for calcium monohydrosulfide (CaSH), calcium monoamide (\ce{CaNH2}), calcium monomethoxide (\ce{CaOCH3}), 
and calcium monophenoxide (CaOPh) using the cc-pwCVTZ basis set for Ca and cc-pVTZ basis sets for S, C, N, O, and H. 
The Ca 3s, 3p, and 4s electrons together with the valence electrons in S, C, N, O, H have been correlated 
in the CC calculations. 
The dipole derivatives have been obtained by means of numerical differentiation of dipole moments using a two-point formula and
a stepsize of $0.1Q$. The use of a step size of $0.01Q$ gives essentially the same results. 
The calculated blackbody excitation induced lifetimes at 300 K for the vibronic ground state
of CaSH, \ce{CaNH2}, \ce{CaOCH3}, and CaOPh are shown in Tab. \ref{tab:largerlifetimes}
and compared with the corresponding results for CaF, CaOH, SrOH, and YbOH. 
Based on the benchmark calculations for CaOH in Sec. \ref{sec:abinitio}.B, the errors in the computed lifetimes
are estimated to be around 20\%. 

\subsection{Results and discussion}\label{polyrd}

\begin{table}[]
    \centering
    \begin{tabular}{l|c|c|c}
    \hline
    \hline
    Molecule & Lifetime (s) & $N_\text{vib}$ & $N(\tau_i < 20\text{ s})$ \\
    \hline
    \ce{CaF} & {4.0} & 1 & 1 \\
    \ce{CaOH} & 1.1 & 4 & 3 \\
    \ce{SrOH} & 1.3 & 4 & 3 \\
    \ce{YbOH} & 1.2 & 4 & 3 \\
    \ce{CaSH} & 3.7 & 3 & 2 \\
    \ce{CaNH2} & {1.7} & 6 & {2} \\
    \ce{CaOCH3} & {2.5} & 12 & 2 \\
    \ce{CaOPh} & 1.2 & 33 & 6 \\
    \hline
    \hline
    \end{tabular}
    \caption{Calculated blackbody lifetimes at 300K for laser-coolable molecules in the harmonic approximation. Also shown are the total number of vibrational modes, $N_\text{vib}$, and the number of modes with blackbody excitation times less than 20~s (see text).}
    \label{tab:largerlifetimes}
\end{table}

As shown in Tab. \ref{tab:largerlifetimes}, 
the 300 K blackbody lifetimes for CaSH, \ce{CaNH2}, \ce{CaOCH3}, and CaOPh are consistently $>$1 s, despite the increasing number of vibrational modes, $N_\text{vib}$, in these molecules. Of the molecules studied here, the linear alkaline earth monohydroxides \ce{CaOH}, \ce{SrOH}, and \ce{YbOH}, and the largest molecule, \ce{CaOPh}, have the shortest ground state blackbody lifetimes. Interestingly, \ce{CaOPh} is calculated to have a similar blackbody lifetime to \ce{CaOH} despite containing nearly $10\times$ more vibrational modes.

To explore these findings further, we consider the contribution of each individual vibrational mode to the total calculated blackbody lifetime. The ``partial'' lifetime of the ground state due to blackbody excitation to the $i$th vibrational mode is
\begin{equation}
    \tau_i = \left(\Gamma_{\text{BBR},i}\right)^{-1} = \left(\frac{4\omega_i^3  |\vec{\mu}_{v,i}|^2}{3c^3}\frac{1}{e^{\omega_{i}/k_B T}-1}\right)^{-1} \label{taui}
\end{equation}
where $\omega_i$ is the energy of the $i$th mode and $|\vec{\mu}_{v,i}|$ is the transition dipole moment. Given the overall lifetimes of $\sim$1 s for the molecules considered here, we choose a cutoff of $\tau_i<20$ s to identify vibrational modes that significantly contribute to the blackbody lifetime of the molecule. The number of vibrational modes, $N(\tau_i<20\text{ s})$, that fulfill this criterion is tabulated in Tab. \ref{tab:largerlifetimes}. Tab. \ref{tab:lifetimebreakdown} lists all such vibrational modes, along with their energies, transition dipole moments, and partial lifetimes $\tau_i$, for each of the \ce{Ca}-containing molecules considered in Tab. \ref{tab:largerlifetimes}.

\begin{table}[]
    \centering
    \begin{tabular}{ll|ccc}
    \hline
    \hline
    Molecule & Mode & Frequency & $|\vec{\mu}_v|$ & $\tau_i$ (300K) \\
    \hline
    \ce{CaF} & \ce{Ca-F} stretch & 577 & 0.25 & {4.0} \\
    & Total  & -- & -- & {4.0} \\
    \hline
    \ce{CaOH} & Bend ($d=2$) & 381 & 0.31 & 1.6 \\
    & \ce{Ca-O} stretch & 625 & 0.25 & 3.8 \\
    & Total & -- & -- & 1.1 \\
    \hline
    \ce{CaSH} & \ce{Ca-S} stretch & 311 & 0.17 & 13.2 \\
    & Bend & 364 & 0.25 & 5.1 \\
    & Total  & -- & -- & 3.7 \\
    \hline
    \ce{CaNH2} & {\ce{NH2} out-of-plane bend} & 453 & 0.31 & 2.9 \\
     & \ce{Ca-N} stretch & 545 & 0.22 & 5.0 \\
    & Total  & -- & -- & 1.7 \\
    \hline
    \ce{CaOCH3} & \ce{Ca-O} stretch & 486 & 0.21 & 6.1 \\
    & \ce{C-O} stretch & 1222 & 0.34 & 5.3 \\
    & Total  & -- & -- & 2.5 \\
    \hline
    CaOPh & \ce{Ca-O} stretch & 309 & 0.20 & 9.6 \\
    & ring stretch & 629 & 0.15 & 10.6 \\
    & CH out-of-plane bend & 778 & 0.17 & 9.6 \\
    & ring stretch & 899 & 0.23 & 5.9 \\
    & \ce{C-O} stretch & 1355 & 0.39 & 5.5 \\
    & ring stretch & 1552 & 0.29 & 17.3 \\
    & Total  & -- & -- & 1.2 \\
    \hline
    \hline
    \end{tabular}
    \caption{Breakdown of blackbody lifetimes of Ca-containing molecules by modes with significant contribution.
    The lifetimes $\tau_i$'s at 300K (in s) were calculated in the harmonic approximation using the computed harmonic vibrational frequencies (in cm$^{-1}$) and transition dipole moments $|\vec{\mu}_v|$'s (in debye).}
    \label{tab:lifetimebreakdown}
\end{table}

From the calculations it is apparent that, for the class of alkaline earth--ligand radicals explored here, the number of ``blackbody-active'' vibrational modes remains unchanged or only slowly increases as a function of the total number of modes. While 75\% (3 out of the 4) of vibrational modes in \ce{CaOH} make significant contributions to its blackbody lifetime, in \ce{CaOCH3} and \ce{CaOPh} the numbers drop to 17\% (2 of 12) and 18\% (6 of 33), respectively.
The remaining vibrational modes make only negligible contributions to blackbody excitation because they either have small derivatives $|\vec{\mu}_v| \propto d\vec{\mu}_e/dQ$ or have vibrational frequencies far from the peak of the blackbody spectrum (or both).

Note that the M-O stretching and M-O-H bending modes in CaOH, SrOH, and YbOH are perfectly positioned to be blackbody-excited, thereby causing short ground-state lifetimes. These vibrations each have a large value of $|\vec{\mu}_v| \propto d\vec{\mu}_e/dQ$, i.e. vibrational excitation induces substantial changes in the dipole moment. At the same time, the vibrational frequencies of these modes are not too much higher than $k_BT$ (around 210 cm$^{-1}$ at room temperature), so the exponential factor $\frac{1}{e^{\omega_{i}/k_B T}-1}$ in Eq. \ref{taui} is not too small. 
In contrast, most vibrations in the organic functional groups considered here
do not significantly contribute to the blackbody lifetimes. 
The low-frequency vibrations in these functional groups due to torsional motions
only make small contributions because of their low transition energies $\omega_i$, which are well below the peak of the blackbody spectrum. In addition, many of these vibrations (e.g. rocking and breathing modes within the methyl group in \ce{CaOCH3} or the benzene ring in \ce{CaOPh}) do not substantially alter the length of the molecule along the axis $\vec{R}$ aligned with the molecule-frame dipole moment. The stretching modes in the functional groups, which significantly perturb the length of the molecule along the dipole axis $\vec{R}$, and therefore the electric dipole moment $\vec{\mu}_e\sim e\vec{R}$, indeed have significant vibrational transition strengths $|\vec{\mu}_v| \propto d\vec{\mu}_e/dQ$. However, many of these modes have vibrational frequencies substantially higher than $k_BT$. Their contributions to blackbody radiation induced lifetimes thus are quenched by the exponential factor in Eq. \ref{taui}.

Taking \ce{CaOCH3} as an example, Tab. \ref{tab:lifetimebreakdown} indicates that the most significant contributions to the blackbody lifetime come from the \ce{Ca-O} stretch and \ce{O-C} stretch modes.
Each of these modes substantially changes the length of the molecule along the principal axis (see the supplementary material to Ref. \cite{mitra2020direct} for an illustration). Following this intuition, the Ca-O-C bending mode at 151 cm$^{-1}$ and the \ce{CH3} stretching mode at {2985 cm$^{-1}$} also have significant vibrational transition moments of $|\vec{\mu}_v| = 0.14$ and 0.15~D, respectively. However, blackbody excitation to these states is suppressed by the small power spectral density of 300~K blackbody radiation at these frequencies.

While the arguments above provide a coarse intuition for the ground state blackbody lifetimes of polyatomic molecules considered here, calculations will ultimately be required to determine the lifetimes for other molecules and molecular structures. Nonetheless, this work provides preliminary evidence that lifetimes no shorter than that of \ce{CaOH} can be expected for a number of complex laser coolable polyatomic molecules. Blackbody lifetimes can be significantly extended for all molecules by cooling the surrounding environment to cryogenic temperatures.

\section{Conclusion}
\label{sec:conclusion}

In summary, we have developed rate equations to model the rovibrational thermalization dynamics of optically trapped CaOH molecules and described a fitting procedure to measure radiative and blackbody lifetimes \cite{hallas2022optical}. We have calculated the lifetimes using \emph{ab initio} theory and find good agreement with the experimental results. The same calculations are used to predict radiative and blackbody lifetimes for the molecules \ce{SrOH} and \ce{YbOH}, which have similar structure to \ce{CaOH}. Finally, we have performed \emph{ab initio} calculations of ground-state blackbody lifetimes for larger polyatomic molecules that appear amenable to laser cooling, finding that these lifetimes are $>$1~s for a number of different structures despite the large number of vibrational modes. In all cases, blackbody lifetimes can be dramatically increased by cooling the environment to cryogenic temperatures.

Taken together, this study should inform current and future experiments with trapped, quantum-state-controlled polyatomic molecules. The radiative lifetimes measured here set a limit on the interrogation time for experiments making use of parity-doublet structure in excited vibrational bending modes of linear triatomic molecules \cite{kozyryev2017precision, augustovivcova2019ultracold, kozyryev2021enhanced, anderegg2023quantum}. For other polyatomic molecules where parity-doublet structure can be found in the ground vibrational state \cite{wall2013simulating, wall2015realizing, yu2019scalable, augenbraun2020molecular}, blackbody excitation may be an experimental limitation over timescales of $\sim$1~s, though can be effectively eliminated by cooling the surrounding environment, enabling a route to long coherence times for quantum science applications using the complex degrees of freedom of trapped polyatomic molecules.

\begin{acknowledgments}
We thank Benjamin Augenbraun and Ashwin Singh for useful discussions. This work was supported by the AFOSR and the NSF. NBV acknowledges support from the NDSEG fellowship, LA from the HQI, and PR from the NSF GRFP. The computational work at the Johns Hopkins University was supported by the National Science Foundation under Grant No. PHY-2011794.
\end{acknowledgments}

\bibliographystyle{apsrev4-2}
\bibliography{CaOHReferences}

\end{document}